\documentclass[useAMS,usenatbib]{aastex61}
\usepackage{gensymb}
\usepackage{hyperref}
\usepackage{graphicx}
\usepackage{longtable}
\usepackage{lipsum}
\usepackage{relsize}
\usepackage{textcomp}
\usepackage{amsmath}

\shorttitle{Asteroid 2014 JO25}
\shortauthors{Kumar V. et al.}

\begin{document}

\title{\hspace*{-0.06in}Time and phase resolved optical spectra of potentially hazardous asteroid 2014 JO25}

\correspondingauthor{Kumar Venkataramani} 
\email{kumar@prl.res.in} 

\correspondingauthor{Shashikiran Ganesh}
\email{shashi@prl.res.in}


\author{Kumar Venkataramani}
\affiliation{Astronomy \& Astrophysics Division, Physical Research Laboratory, Navarangpura, Ahmedabad}
\affiliation{Indian Institue of Technology, Gandhinagar }

\author{Shashikiran Ganesh}
\affiliation{Astronomy \& Astrophysics Division, Physical Research Laboratory, Navarangpura, Ahmedabad}

\author{Archita Rai}
\affiliation{Astronomy \& Astrophysics Division, Physical Research Laboratory, Navarangpura, Ahmedabad}
\affiliation{Indian Institue of Technology, Gandhinagar }

\author{Marek Hus\'{a}rik}
\affiliation{Astronomical Institute of the Slovak Academy of Sciences, SK-05960 Tatransk\'{a} Lomnica, Slovakia}


\author{K.S.Baliyan}
\affiliation{Astronomy \& Astrophysics Division, Physical Research Laboratory, Navarangpura, Ahmedabad}

\author{U.C.Joshi}
\affiliation{Astronomy \& Astrophysics Division, Physical Research Laboratory, Navarangpura, Ahmedabad}

\label{firstpage}

\begin{abstract}

The asteroid 2014 JO25, considered to be ``potentially hazardous" by the Minor Planet Center, was spectroscopically followed during its close-Earth encounter on 19th and 20th of April 2017. The spectra of the asteroid were taken with the low resolution spectrograph (LISA), mounted on the 1.2-m telescope at the Mount Abu Infrared Observatory, India. Coming from a region close to the Hungaria population of asteroids, this asteroid follows a comet-like orbit with a relatively high inclination and large eccentricity.  Hence, we carried out optical spectroscopic observations of the asteroid to look for comet-like molecular emissions or outbursts. However, the asteroid showed a featureless spectrum, devoid of any comet-like features.   

The asteroid's light curve was analyzed using V band magnitudes derived from the spectra and the most likely solution for the rotation of the asteroid was obtained. The absolute magnitude $H$ and the slope parameter $G$ were determined for the asteroid in V filter band using the IAU accepted standard two parameter H-G model. A peculiar, rarely found result from these observations is its phase bluing trend. The relative B-V color index seems to decrease with increasing phase angle, which indicates a phase bluing trend. Such trends have seldom been reported in literature. However, phase reddening in asteroids is very common. The asymmetry parameter $g$ and the single scattering albedo $w$ were estimated for the asteroid by fitting the Hapke phase function to the observed data. The asteroid shows relatively large value for the single scattering albedo and a highly back scattering surface.
 
\end{abstract}

\keywords{Minor planets, asteroids: individual(2014 JO25) - methods: observational - techniques: spectroscopic - telescopes}

\section{Introduction}

Asteroids are a class of minor bodies of the solar system which move in different orbits around the Sun. Out of the number of asteroids discovered till date, some could either be pristine material formed from the early stages of the solar nebula or they could also be fragments formed from the impacts and collisions of larger bodies and failed planets perturbed by the gravitational influence of the giant planets like Jupiter and Saturn. A major chunk of these bodies are found in the main-belt  lying between Mars and Jupiter, at heliocentric distances between 2.1 and 3.3 AU \citep{PS}. All those bodies which fall into the Kirkwood gaps in this zone come under strong influence of Jupiter and are thrown into the inner-solar system. This forms one of the sources of the near Earth objects. All  asteroids and comets having a perihelion distance less than 1.3 AU are classified into the near Earth population. Another major source for NEO's is the fragmented parts resulting from impacts and collisions occurring in the main belt.   By looking into the surface properties of these broken chunks, we can expect to study the interiors of the parent asteroid \citep{asteroid_chunks}.

Many surveys have been performed for the near Earth population. The Near-Earth Asteroid Tracking System \citep{NEAT} operating autonomously at Maui Space Surveillance Site, Hawaii, has discovered many near Earth Objects (NEO's). Some of the other near Earth asteroid surveys are Catalina Sky Survey \citep{Catalina}, Lincoln Near Earth Asteroid program \citep{LINCOLN}, Pan-STARRS asteroid survey \citep{panstarrs1, panstarrs2}, the Infra-red space based observatory survey NEOWISE \citep{neowise}. All of them have contributed immensely to the rapid discovery of a large number of near Earth asteroids and comets.

The asteroid 2014 JO25 (JO25 for all future text references in this paper) is one such near Earth asteroid belonging to the Hungaria family of asteroids. The asteroid was discovered in May 2014 by A. D. Grauer in the Catalina Sky Survey as a part of NASA's NEO observation program. The asteroid had a very close Earth flyby on 19th April 2017. At the closest point, the asteroid was within 4.6 lunar distance\footnote{JPL NASA online ephemeris (\url{https://ssd.jpl.nasa.gov/horizons.cgi})} of Earth. According to JPL Horizons database, the asteroid has a semi-major axis of 2.06 AU, eccentricity of 0.88, perihelion distance of 0.237 AU and an orbital period of 2.97 years. The asteroid has been confirmed to be a double lobed contact binary\footnote{Images obtained from Goldstone Solar System Radar as a part of Goldstone Deep Space Network Mission (\url{https://echo.jpl.nasa.gov/asteroids/2014JO25/2014JO25_planning.html})} \citep{goldstone_radar}. 

The Tisserand parameter (T$_J$) is used to distinguish asteroid and comet orbits with reference to the Sun and Jupiter as the major influencing bodies. 
For most of the asteroids T$_{J}$ $>$ 3 and those objects showing a comet-like behaviour have T$_{J}$ $<$ 3. There are some comets (Encke type) which also tend to have T$_{J}$ $>$ 3.  \citet{DC} have studied the near Earth asteroid population and some of the dormant comets which are characterised by 2 $<$ T$_{J}$ $<$ 3. They have found that half of the near Earth asteroids on a comet-like orbit would exhibit comet-like albedos. For the asteroid JO25, T$_{J}$ $=$ 3.04 with a relatively high orbital inclination of 25$^{\circ}$ to the ecliptic. These values imply a comet-like orbit for this asteroid. This gives a cue to look for comet-like emissions in its optical spectrum. With this in mind, we have carried out spectroscopic follow up of JO25 at varying solar phase angles and determined its colour variation, absolute magnitude, rotation period, the Hapke parameters with a few approximations, and the spectral reflectance.  This enables us to make inferences on the surface properties and behaviour of the asteroid. 


\section{Observations and Data Analysis}

Spectroscopic observations were made using the LISA spectrograph with 1.2-m telescope at the Mount Abu Infrared Observatory, Mount Abu, India. LISA\footnote{More details on the spectrograph are available on the web site of the manufacturer: Shelyak Instruments (\url{http://www.shelyak.com/rubrique.php?id_rubrique=12&lang=2})} is a low resolution long-slit based spectrograph designed for the study of faint and extended objects. Details and specifications of the instrument have been described in \citet{c2014q2}. The typical wavelength coverage is from $\scriptstyle\mathtt{\sim}$ 3800 \AA $ $  to $\scriptstyle\mathtt{\sim}$ 7400 \AA $ $  with a wavelength scale of 2.6 \AA $ $ per pixel. The telescope has a focal ratio of f/13 while the instrument has been designed for a focal ratio of f/5 to f/8. Therefore a 0.5x focal reducer was used to match the focal ratio of the instrument with that of the telescope. This results in a  plate scale of 0.25 arcsec per pixel on the LISA CCD plane and the 35$\mu$m slit covers 1.75 arcsec.

LISA was mounted on the 1.2-m telescope on 19th and 20th of April 2017 to follow the flyby of the asteroid JO25. On 19th April, we covered a solar phase angle ranging from 66$^{\circ}$ to 48$^{\circ}$, whereas on 20th, the phase angle remained close to 26$^{\circ}$. The asteroid was tracked using a variable track mode of the telescope which allowed to follow the non-sidereal movement of the asteroid during the entire period of observation. Manual guiding corrections were also required, occasionally, to keep the asteroid fixed in one of the pixel. The telescope pointing was adjusted in such a way that the asteroid was positioned at the center of the slit.

On 19th April 2017, the asteroid JO25 was spectroscopically monitored for a total period of 5.76 hrs (16:15:10 UT to 22:01:34 UT). There was an intermediate time gap of 1.71 hrs, which was utilised for observing the spectrophotometric standard stars and the solar analog stars. The first part of the monitoring was done for 2.38 hrs and second part for a period of 1.67 hrs. Therefore we obtained a total of 40 frames with each spectrum having an exposure time of 400 seconds. On 20th, the asteroid was covered for a continuous period of 4.59 hours (17:59:16 UT to 22:34:28 UT).
Spectrophotometric standard stars were observed on each night of observation for the purpose of flux calibration and airmass/extinction correction. Apart from these, solar analog (G2V type) stars were also observed in order to obtain the reflectance of the asteroid.

\begin{figure*}
  \centering
  \hspace*{-0.4in}
   \includegraphics[width=1.1\textwidth]{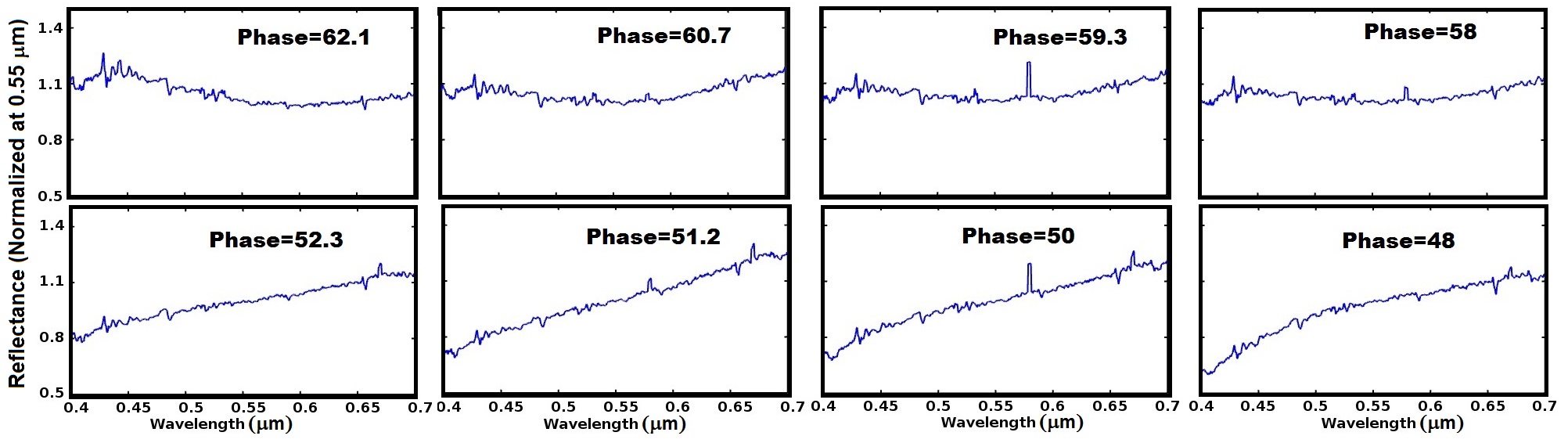}
   \caption{Solar phase angle resolved reflectance spectra of the near Earth Asteroid 2014 JO25.}
 \label{spectra}
\end{figure*}

The spectral data were reduced by following the standard procedures of trimming, bias subtraction, flat-fielding, cosmic ray correction, extraction of one dimensional spectra from the raw image, wavelength calibration, extinction correction and flux calibration. All these were carried out using the spectral reduction package available in Image Reduction and Analysis Facility (IRAF 2.16a). The APALL task in IRAF was used for defining the aperture for the extraction of spectra, tracing the aperture along the dispersion axis and for background subtraction. Background sky correction was applied by placing apertures at various locations along the slit and fitting a polynomial function to determine the extent of background level before subtraction. Argon-Neon lamp spectra were obtained at regular intervals during each night of observation. These were used for the purpose of wavelength calibration. Some of the standard stars observed for calibration purpose were HD74721 and BD33d2642.  All the spectra including those of the standards were corrected for airmass and extinction. The instrument sensitivity function was then obtained by fitting a spline function over the ratio of number of counts in the spectra of the star to the absolute flux value obtained from the catalog. This function was then used to flux calibrate a known star and obtain its integrated synthetic magnitude (section \ref{SM}). The fit was varied, till we obtained the best numbers for the B,V and R magnitudes for the star and its B-V, V-R colours as compared to the expected values\footnote{Values obtained from \url{http://simbad.u-strasbg.fr/simbad/}}. The resulting sensitivity function was then used for flux calibration of the asteroid.

In order to obtain the reflectance spectra of the asteroid, various G2 type stars were observed. Each of these spectra were checked for their absorption line features, as compared to the solar absorption features seen in the spectra of the asteroid. Two solar analog stars HD140514 and HD126991, which have been regularly observed, were found to best match the absorption features. Therefore, these two stars were selected and the flux levels were scaled. The asteroid spectra were divided by the spectra of these solar analog stars to obtain the reflectance of the asteroid. The normalized reflectance spectra of the asteroid 2014 JO25 at various solar phase angles is shown in the figure \ref{spectra}.

\subsection{Magnitudes derived from the spectra}
\label{SM}
The B-V-R magnitudes and colours were extracted from the asteroid spectra at different phase angles. The magnitudes were obtained in the following way (also partly described by \citet{c2014q2} for Af$\rho$ calculations): \\
The transmission profiles of the Johnson-Cousin UBVRI filters were measured using a UV-Visible spectrometer with the wavelength step sizes of 1 \AA $ $ and 10 \AA $ $. These measured profiles were re-sampled to match the resolution of the asteroid spectra. The synthetic magnitudes were then calculated as 
\begin{equation}
 F_{i}(\lambda)=\frac{\int f(\lambda)T_{i}(\lambda) d\lambda}{\int T_{i}(\lambda) d\lambda }
\end{equation}
where $f(\lambda)$ is the flux from the observed asteroid spectra (in flux units obtained from flux calibrations), $T_{i}(\lambda)$ is the transmission profile (in percentage) of the $i^{th}$ filter and $F_{i}(\lambda)$ is the total integrated flux in the $i^{th}$ filter. The integrated flux was then converted to corresponding magnitude scales. 
We have confirmed that the results from this method are in agreement with the magnitudes of the known spectrophotometric standards.

\subsection{The H-G magnitude system : Phase curves}
\label{H-G-theory}
The two parameter H-G function developed by \citet{bowell} has been universally accepted as the standard way of expressing the absolute magnitudes of the solar system bodies. It takes into account the change in visual magnitude of a body with respect to the changing solar phase angle.  The magnitude-phase curve for each body provides significant information about the surface structure and behavior. The model is developed based on the scattering theories of \citet{lumme-bowell} and has two parameters to describe the phase function. One is the absolute magnitude H and the other is the slope parameter G. The two parameter equation goes as follows.
\begin{equation}
\label{HG_equation}
 H=H(\alpha)+2.5 \log[(1-G)\phi_{1}(\alpha)+G\phi_{2}(\alpha)]
\end{equation}
where $\phi_{i}=\exp(-A_{i}(\tan(\alpha/2)^{B_{i}}))$, $i=1,2$. The constants $A_{1}=3.33$, $A_{2}=1.87$, $B_{1}=0.63$ and $B_{2}=1.22$ are taken from \citet{bowell}.

\subsection{The Hapke phase function model}

\label{hapke_sec}
\citet{hapke_1} first proposed a model to predict the physical characteristics of a planetary body based on its scattering properties. It then went on to become a rigorous photometric model, used to estimate the physical properties of the regolith from distant planetary bodies. Hapke started by deriving an approximate analytic solution to the radiative transfer equation which describes the light scattering principles from particulate surfaces. He has given analytical expressions for bidirectional reflectance, integral phase functions and a few other physical quantities. The integral phase function is a distribution function of the amount of light being scattered into a particular direction from the total integrated disk brightness. The single particle phase function is the amount of scattered light from a single particle. There have been different approaches to calculate the single particle phase function $p(\alpha)$. It can be represented as a linear combination of legendre polynomials \citep{hapke_book}. 
\begin{equation}
 p(\alpha)=\sum_{i=0}^{\infty} b_{i} p_{i}(\alpha)
\end{equation}

\citet{hapke_1,hapke_2,hapke_3} used the first order expansion of the phase function, whereas most of the real functions would have been best estimated by a second order expansion. \citet{hapke_1} had used the Lommel-Seeliger law in order to estimate the radiance scattered from the surface and thereby derive the integral phase function.  \citet{hapke_2, hapke_3} estimated the correction factors for the macroscopic roughness ($\theta$), the extinction co-efficient and the opposition effect.  Yet another form of the single particle phase function is the one used by \citet{buratti}. She used the single Henyey-Greenstien function to represent the $p(\alpha)$, which is given as 
\begin{equation}
 p(\alpha)=\frac{1-g^{2}}{(1+2gcos(\alpha)+g)^{3/2}}
\end{equation}
where $g$ is the asymmetry parameter and can have a value ranging from -1 to 1. A negative value of $g$ indicates back-scattering and a positive value indicates the presence of forward scattering phenomena from the surface of the observed object. 

The final form of the Hapke integral phase function that we have followed in our calculations is as given by equation 21 in  \citet{bowell}. For the sake of completeness, we include the final equation, which is given as:

\begin{equation}
\begin{split}
\label{hapke_eq}
 \Phi(\alpha)=&\frac{K(\alpha,\theta)}{A_{p}} \\ &\left[\left \{ \frac{w}{8}\left[(1+B(\alpha))p(\alpha)-1\right]+\frac{r_{o}}{2}(1-r_{o}) \right\}  
      \left\{1-sin\left(\frac{\alpha}{2}\right) tan\left(\frac{\alpha}{2}\right)    ln\left(cot\left(\frac{\alpha}{4}\right)\right)          \right\} + \frac{2r_{o}^{2}}{3\pi} [sin(\alpha) + (\pi - \alpha)cos(\alpha)]  \right]
\end{split}
\end{equation}

where $A_{p}$ is the physical albedo of the scattering body, given as

\begin{equation}
 A_{p}=\frac{w}{8}[(1+B_{0})p(0)-1]+\left(\frac{r_{0}}{2}+\frac{r_{0}^{2}}{6} \right)
\end{equation}

the expressions for roughness correction factor $K(\alpha,\theta)$, opposition effect function $B(\alpha)$, and the bihemispherical reflectance $r_{0}$ are given in \citet{bowell}. $B_{0}$ is the amplitude of the opposition effect. The phase function $\Phi(\alpha)$ was substituted for the term $(1-G)\phi_{1}(\alpha)+G\phi_{2}(\alpha)$ in the equation \ref{HG_equation} in order to calculate the reduced magnitude and fit it to the observed data.

With the use of single Henyey-Greenstien particle phase function, the Hapke photometric model has five unknown parameters. They are (i) Single Scattering Albedo `$w$', (ii) Asymmetry Parameter `$g$', (iii) Macroscopic roughness parameter `$\theta$', (iv) Amplitude of opposition effect `$B_{0}$' and (v) The angular width of the opposition effect `$h$'. Of these, the opposition effect amplitude and angular width can be determined with accuracy only if we have a coverage of very low phase angles (atleast $\alpha$ $<$ 3$^{\circ}$). The roughness parameter can be confidently estimated at higher phase angles ($\alpha$ $>$ 120$^{\circ}$). The asymmetry parameter ($g$) and the single scattering albedo ($w$) can be estimated at intermediate phase angles.

\section{Results and Discussion}
The solar phase angle of JO25 varied to large extent on 19th April 2017 as compared to the next day. Therefore, only the data of the first day were utilized in order to study and model the physical characteristics of the asteroid surface resulting from the phase change. The asteroid spectrum was almost featureless, except for some minor absorption features around 0.49 $\mu$m, 0.55 $\mu$m  and 0.60 $\mu$m. These features varied with the phase angle. However, they were too weak to characterize the asteroid into a particular taxonomic class.
From the spectra we obtained the V band light curve, and the B-V color of the asteroid using the Johnson's filter profile for B and V filters. The apparent visual magnitude (derived from the spectra) ranges from 10.70 $\pm$ 0.01  to 12.31 $\pm$ 0.01 during the period of our observations on 19th April. We then obtained the absolute magnitude and the slope parameter using the two-parameter H-G phase function. We also modeled the phase function using Hapke's theory, by estimating some of the key parameters like the single scattering albedo, the asymmetry parameter and the macroscopic roughness of the asteroid surface.

\subsection{The phenomena of phase bluing }
\begin{figure*}
\centering
  \includegraphics[width=0.8\paperwidth,height=0.6\textwidth]{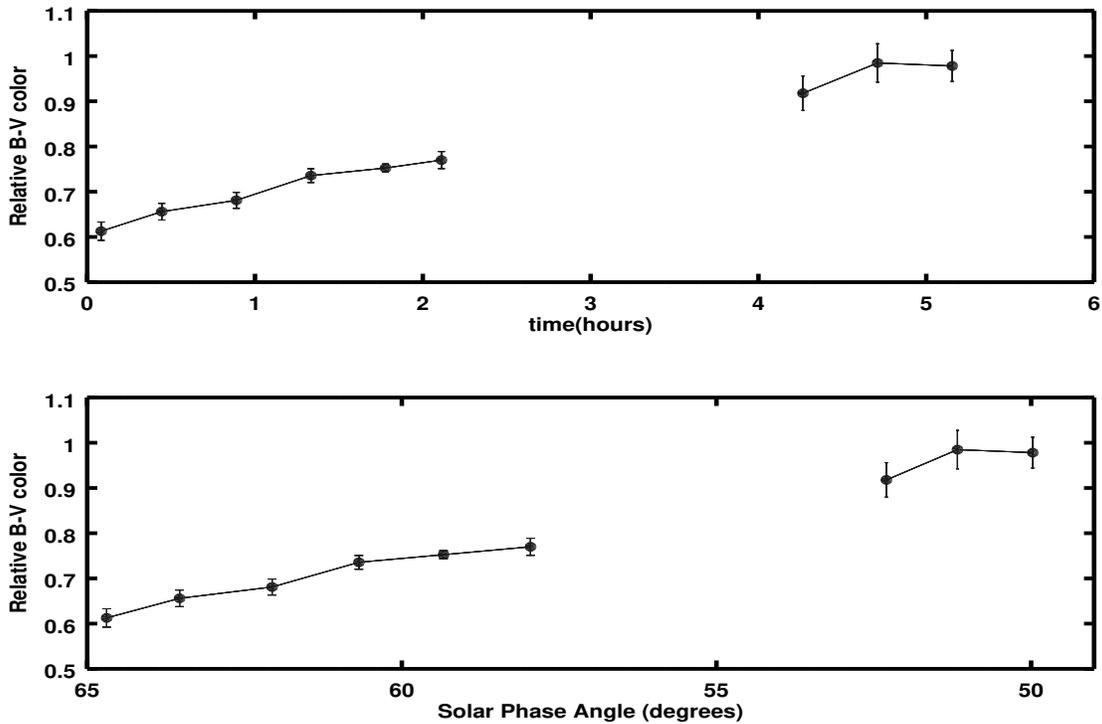}
 \caption{The B-V color of the asteroid as a function of time (top panel) and solar phase angle (bottom panel). The asteroid seems to be more blue at higher phase angles. }
 \label{asteroid_BV}
\end{figure*}

The V-R color which corresponds to the red side of the spectrum turns out to be almost constant, whereas the blue side of the spectrum changed significantly. The B-V color increased from 0.61 $\pm$ 0.02 to 0.98 $\pm$ 0.03 while the V-R color index remained almost stable at 0.41 $\pm$ 0.07. The errors are calculated as the standard deviation of four consecutive measurements. The plot of B-V color index as a function of solar phase angle and observing epoch (time) is shown in the figure \ref{asteroid_BV}. As seen in the figure,  the asteroid color shows a steep bluing trend with increasing phase angle. As the asteroid approaches its closest point to Earth, the phase angle decreases with time. Thus this asteroid shows a bluer spectra at higher phase angles, contrary to what we expect from most of the asteroids or any other near Earth object. We claim this bluing trend of the color index with a very high degree of certainty, as this is seen in the relative color index. \\      
A majority of the minor bodies, moons and even planets are known to show phase reddening of their spectra. This was first observed by \citet{gradie} in their laboratory experiments, where the wavelength dependence of the phase co-efficients are considered as the major reason for such reddening. Thereafter, phase reddening effects have been seen in many asteroids. \citet{lumme-bowell} have shown such effects using their photometric studies. Phase reddening in near Earth asteroids was first reported by \citet{luu-jewitt}, as their spectral slopes were much redder than those measured for 3:1 resonance asteroids. More recently \citet{sanchez} have investigated the effect of phase reddening on the spectroscopic behavior of the near Earth asteroids and found that ordinary chondrite spectra shows a significant increase in the spectral slope for phase angles greater than 30$^{\circ}$. Reddening also induces large changes in the band depth. This has been seen by \citet{sanchez} in both asteroids and meteorite samples. They also report phase reddening which is comparable to effects caused by space weathering.

As compared to the reddening effect, phase bluing has hardly been reported, except for claims from some space missions and flybys. Phase bluing can be defined as an effect that produces a decrease in the spectral slope with increasing phase angle. One important instance of such a result was from the Viking 1 lander on Mars. \citet{guinness} has reported the bluing of reflectance at phase angles greater than 90$^{\circ}$ from the surface of Mars, as measured by Viking 1. The phase bluing effect is not yet well understood, although laboratory experiments reproduce the arch seen in the data from Viking 1 many times. \citet{schroder} have carried out laboratory experiments more recently to understand the process of phase reddening and bluing. They have considered reflectance of different sized fragments of basalt, granite and limestone, examining the variation of color with the phase angle for different wavelength ratios.
Examining the color versus phase plot of the asteroid JO25, we get a negative slope, which is approximately equal to -0.025 mag/degrees, which is relatively steep. Of all the phase plots from \citet{schroder}, the one for granite particles of size 1000 $\mu$m (figure 17.D) seems to be the steepest.

According to \citet{schroder}, phase reddening might be due to selective absorption, in which the material favors the absorption of blue wavelength and scatters the redder side. In this case of phase bluing, clearly the red wavelengths are being absorbed more than the bluer ones.  Such selective absorption would be dependent on the surface roughness at scales comparable to the wavelength being scattered. \citet{schroder} also comment, that their model explaining the reddening arch may not be a complete one, as the amplitude of the observed arch of granite is larger than the simulated one. 

Space weathering results in changing colors of the surface and reddening of the phase curve in asteroids. The weathered rocks appear redder as compared to the fresh interiors. The study by \citet{age-color} has shown a genetic relationship between the fresh chondrites and the S-complex asteroids. Therefore, there is a chance, that the part of the asteroid JO25 which shows bluer color (at higher phase) may be freshly exposed as compared to the part showing relatively redder color (at lower phase). This suggests that the surface of the asteroid JO25 is highly heterogeneous. However, this strong gradient in the color with the changing phase may also be an illusive viewing effect due to the rapidly changing phase of the asteroid coupled with its double lobed shape. 

Such a phase bluing trend from an asteroid regolith was first reported by \citet{nysa}. They found that the E type asteroid 44 Nysa became bluer with increasing phase (-0.001 to -0.007 mag/deg). In case the of JO25, the phase bluing phenomena shows a large gradient (-0.025 mag/deg). Although the interpretations of phase bluing phenomena are highly debatable, our observations give a strong conclusive evidence for such a trend to exist in nature.

\subsection{The light curve}
\begin{figure*}
  \includegraphics[width=0.95\textwidth]{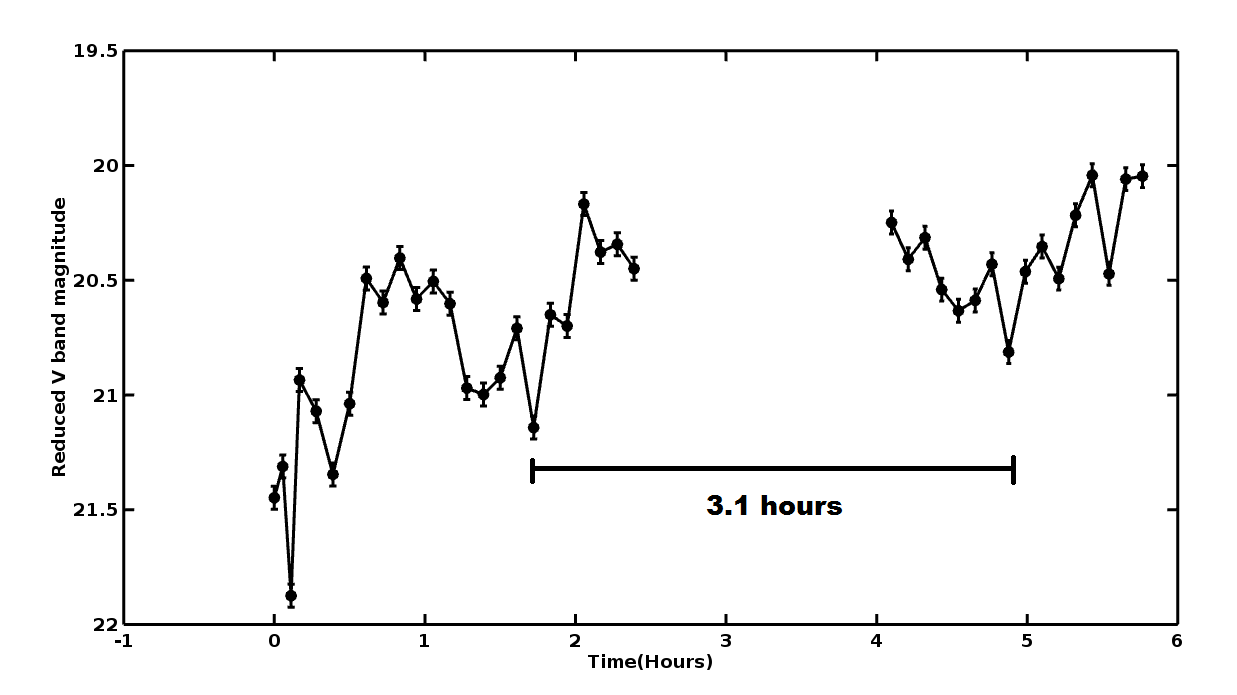}
 \caption{The V band light curve for the asteroid JO25 observed on 19/04/2017.}
 \label{light_curve}
\end{figure*}

\begin{figure*}
  \includegraphics[width=0.95\textwidth]{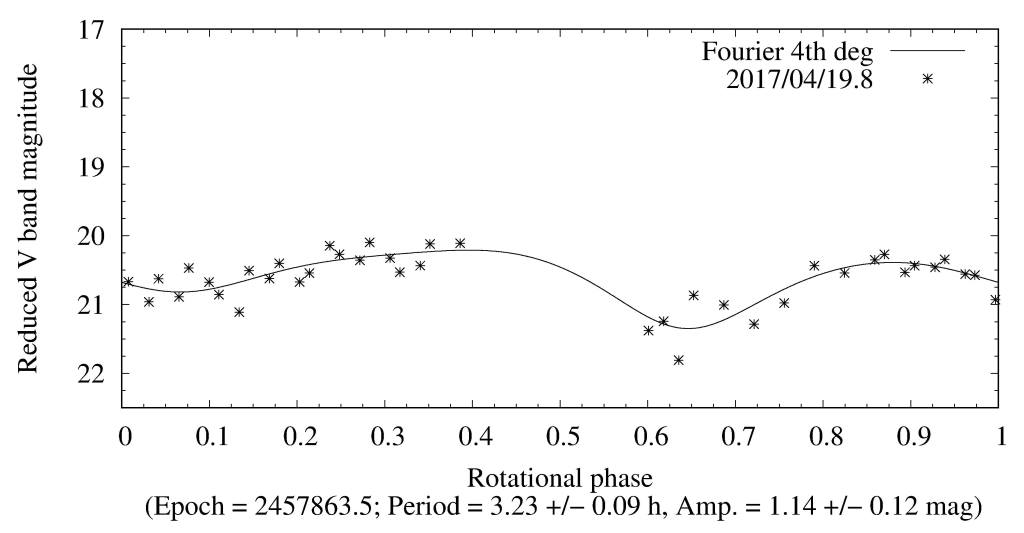}
 \caption{The V band light curve of the asteroid 2014 JO25 observed on 19th April 2017 along with a 4th order Fourier fit to the light curve showing a possible rotation period of 3.23 $\pm$ 0.09 hr. The Y axis shows the reduced magnitude (V band). }
 \label{fourier_1}
\end{figure*}

\begin{figure*}
  \includegraphics[width=0.95\textwidth]{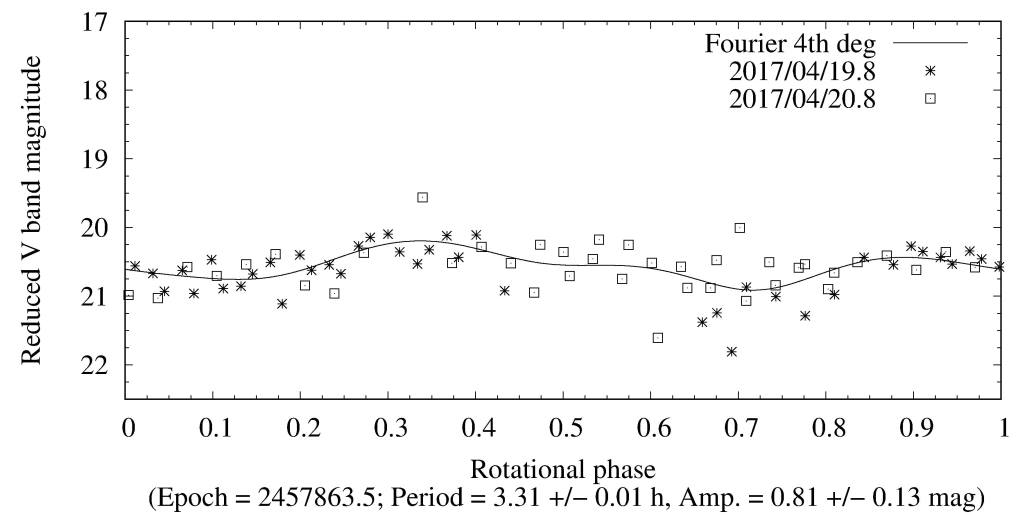}
 \caption{The composite V band light curve of the asteroid 2014 JO25 as observed on 19th and 20th of April 2017 along with a 4th order Fourier fit to the light curve showing a possible rotation period of 3.31 $\pm$ 0.01 hr. The Y axis shows the reduced magnitude (V band).}
 \label{fourier_2}
\end{figure*}

\begin{figure*}
  \includegraphics[width=0.95\textwidth]{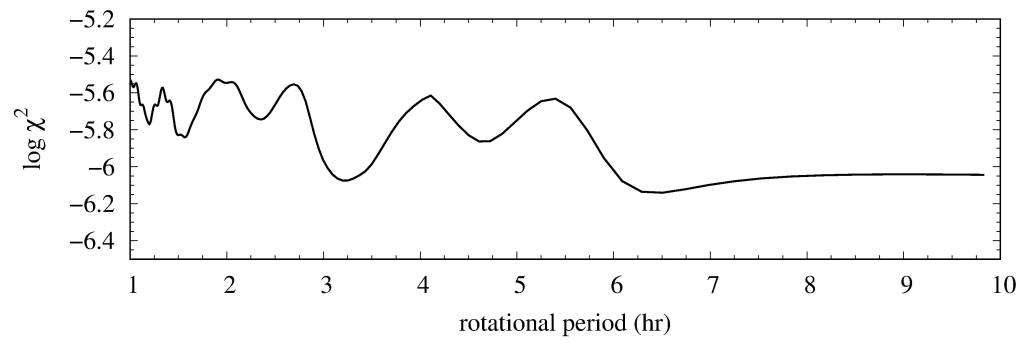}
 \caption{Periodogram for light curve of the asteroid 2014 JO25 as observed on 19th April 2017. The most likely solution for the rotation period of the asteroid is seen to be 3.23 $\pm$ 0.09 hr}
 \label{periodogram_figure}
\end{figure*}

The V band light curve for the asteroid JO25 was obtained from the optical spectra using the synthetic magnitude calculations outlined in section \ref{SM}. The V magnitudes were corrected for heliocentric and geocentric distances and the reduced V magnitude light curve was plotted. This plot is shown in the figure \ref{light_curve}. The light curve is not continuously sampled for the asteroid. The surface of the asteroid being observed also changed over the period of observation. For the initial analysis, we visually identified the pattern which seemed to repeat after a certain span of time from the V band light curve. \\ We estimated the rotation period of the asteroid to be 3.1 hr by noting the difference in the time stamp between these repetitions.
In order to confirm the rotation period, we have done the Fourier analysis of the light curve on both the nights. With a 4th degree Fourier fit, we clearly obtain a solution of 3.2 hr rotation period on 19th April with an amplitude of 1.14 mag. Whereas, one of the best fit solutions of the composite light curve (of both nights) is 3.3 hr with an amplitude of 0.81 mag. The light curve on 19th April and the composite light curve along with the Fourier fit is shown in figures \ref{fourier_1} and \ref{fourier_2}. The periodogram of the data on 19th also clearly shows a minimum at 3.2 hr, which is shown in figure \ref{periodogram_figure}. \\
\citet{goldstone_radar} have obtained a period of 4.5 hrs, whereas \citet{warner_jo25} has claimed three possible periods (3, 4.5 and 6 hours) for the asteroid. This rotation period of 4.5 hrs has also been confirmed by \citet{jo25_mnras} and \citet{jo25_MPB}. \citet{jo25_mnras} have used an extensive data set of 4540 data points to derive the rotation period. However, all of these observations have been made starting from 20th April, when the asteroid's phase was lower than on 19th April when we had observed. Our observations indicate, that along with the 4.5 hr rotation period of the asteroid, a rotation period of $\approx$ 3.2 hr is also a very likely solution, when the asteroid was observed at higher phase angles on 19th April 2017.
\\
We see a large change in the amplitude of the light curve between 19th and 20th April. The amplitude variation of the V band light curve was negligible on 20th April as compared to the previous night. On 19th, the asteroid was at a higher phase angle varying from 66$^{\circ}$ to 48$^{\circ}$, whereas on 20th, it was at a much lower phase of 25$^{\circ}$ and varied only by a degree. \citet{zappala} have studied the amplitude phase relation for asteroids. Their work suggests that the light curve amplitude for an asteroid is very likely to peak at intermediate phase angles.  Our findings are consistent with their results.  On the second night of observations the phase angle remained close to 25$^{\circ}$ and the change in amplitude of the light curve remained small. Another possible reason for the light curve amplitude being small on the second day is the possible orientation of the rotation axis of the asteroid along the line of the sight.

\subsection{The Hapke parameter and H-G two parameter fitting}
\label{hapke_results}
The absolute magnitude $H$ of the asteroid JO25 was calculated using the two parameter H-G model as described in section \ref{H-G-theory}. The parameters $H$ and $G$ were varied simultaneously and the resulting phase function was compared with the observed data (reduced magnitude, independent of heliocentric and geocentric distances). The best fit parameters were obtained by minimizing the chi-square. The absolute magnitude $H$ and the slope parameter $G$ were obtained for the V band filter data. The optimum values obtained are mentioned in table \ref{hapke_results_table}.
\begin{table}
\centering
\caption{ Hapke and H-G parameters (Optimum Estimated Value)  }
\begin{tabular}{|c | c | c | c | c |}
\hline
Filter ($\lambda$ - Wavelength) & \multicolumn{2}{c|}{H-G Model Parameters} & \multicolumn{2}{c|}{Hapke Model Parameters} \\ \cline{2-5}
  & $H$ & $G$ & $g$ & $w$ \\ 
  & (Absolute Magnitude) & (Slope Parameter) & (Asymmetry Factor) & (SSA\footnote{single scattering albedo}) \\ \cline{1-5}
\multicolumn{1}{|c|}{V (5477 \AA) } & 18.6 & 0.20 & -0.39 & 0.25 \\ \hline
\end{tabular}
\label{hapke_results_table}
\end{table}

The phase function for the asteroid was also calculated using the (Hapke) equation \ref{hapke_eq} as given in section \ref{hapke_sec}. We tried using a second order Legendre polynomial approach for calculating the single particle phase function $p(\alpha)$. However, a much better fit to the integral phase function was obtained by using the single Henyey-Greenstien function for  $p(\alpha)$. The physical properties of the asteroid JO25 were examined by estimating the parameters of the Hapke model. Since our data does not span the lower range of phase angles, we set the amplitude of opposition effect $B_{0}$ to be zero. The single scattering albedo ($w$), the asymmetry parameter($g$) and the roughness parameter ($\theta$) were varied simultaneously in order to fit the observed phase function with the model. The best fit parameters were estimated by minimizing the chi-square. However, this resulted in an unreasonably high value of the single scattering albedo. Therefore, we also varied $B_{0}$ simultaneously and obtained an optimum value of $B_{0}$ = 0.65 and $h$ = 0.001.

\begin{figure*}
  \includegraphics[width=0.8\paperwidth,height=0.5\paperwidth]{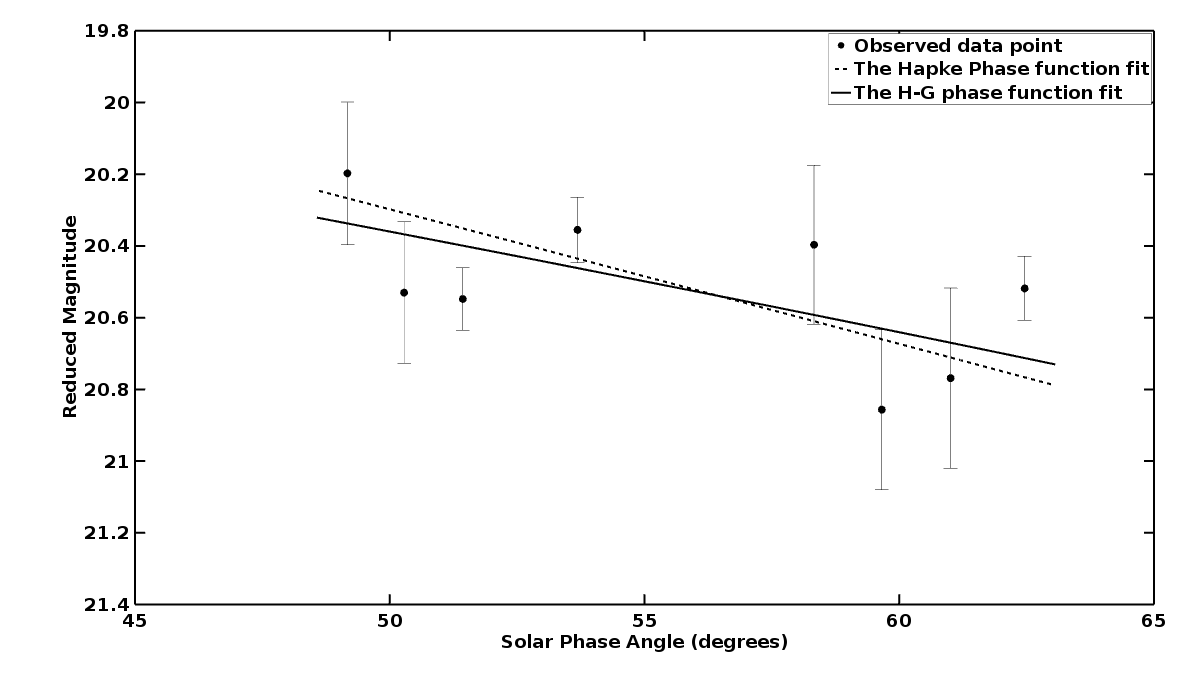}
 \caption{The observed phase function(V band) modelled by estimating the Hapke parameters and the two-parameter H-G phase function }
 \label{hapke-hg-fit}
\end{figure*}

The roughness parameter $\theta$ was varied in steps of 1$^\circ$, whereas the value for the asymmetry parameter $g$ and the single scattering albedo $w$ were varied in steps of 0.01. The parameters were estimated to fit the derived magnitudes obtained using the V band profile. The phase function had almost negligible sensitivity for the entire range of $\theta$ (ranging from 2$^\circ$ to 30$^\circ$). However, we found that the optimum (with minimum chi-square) value  of the roughness parameter was $\theta$ = 14$^{\circ}$. The optimum values obtained for the other two parameters ($g$ and $w$) are given in table \ref{hapke_results_table}.  The best fitting phase function obtained using the two parameter H-G function and the Hapke phase function are shown in the figure \ref{hapke-hg-fit}. 

From the estimated values, JO25 seems to have a relatively moderate albedo and has a back-scattering surface. Interestingly, JO25 is a contact binary asteroid with a shape very similar to that of the comet 67P/Churyumov-Gerasimenko. The moderate value of the single scattering albedo, suggests that the asteroid must fall into an S type (moderate albedo) taxonomic class. The slope parameter in the H-G model turns out to be G=0.2, which also indicates that it belongs to the moderate albedo group \citep{bowell} and falls into the S type taxonomic class. The S type behaviour of the asteroid has been seen by various other studies. The initial results of observations using the SpeX medium resolution spectrograph at the NASA Infrared Telesope Facility (press release - \url{https://cneos.jpl.nasa.gov/news/news196.html}) show a 1 $\mu$m and 2 $\mu$m absorption feature which indicates the S type behavior of the asteroid. \citet{jo25_spectra} have spectroscopically studied this asteroid and they claim that it belongs to S or Q complex of silicaceous asteroids.  \\ 
\newpage

Following are few charecteristics of the asteroid JO25, which is seen from this study:
\begin{itemize}
 \item The asteroid JO25 is extremely bright with relatively moderate value of single scattering albedo. The S type asteroids belong to the moderate albedo group.
 \item The phase bluing phenomenon seen in this asteroid is similar to the one seen in asteroid 44 Nysa \citep{nysa} which is an E type asteroid \citep{Nysa_E_1}.
 \item JO25 shows a featureless spectrum and very steep red slope at smaller phase angles. 
\end{itemize}


The relatively large negative value of the asymmetry parameter $g$ indicates that the surface is highly back-scattering and most of the back-scattering surfaces would be optically thick or opaque. However, if multiple scattering is a reason for the increase in the path length of the light due to which the phase bluing occurs, the optical thickness must also vary substantially over the surface of the asteroid. 

\section{Conclusions}
We have carried out spectral study of the potentially hazardous near earth asteroid JO25 which shows a significant change in its reflectance and color properties, indicating a large amount of heterogeneity in its double lobed surface. A few minor and weak absorption features were noted in the otherwise featureless optical spectra. Using the synthetic magnitude approach to calculate the colors, we obtained the color-phase plot over a significant range of the solar-phase angle. We have obtained the absolute magnitude, slope parameter, single scattering albedo and asymmetry parameter for the asteroid. From our observations and their analysis, we arrive at the following conclusions:

\begin{itemize}
 \item The optical spectrum of the asteroid JO25 does not show any comet-like emissions (though its orbit closely resembles that of an Encke family comet). 
 \item The asteroid shows a strong variation in color with the solar phase angle in the blue side of the spectrum. It shows a bluing trend as the phase angle increases, contrary to the trend seen in most of the planetary, asteroidal or cometary bodies. Such a phase bluing trend has been seen only on another asteroid, 44 Nysa.
 \item The phase bluing trend might indicate the presence of freshly exposed surface, which tends to be more active and volatile as compared to the older and weathered surfaces
 \item It has a highly back-scattering surface and a moderate albedo. This indicates that the asteroid might fall into an S type of asteroid taxonomic class.
 \item Considering all our observational results, the surface of the asteroid JO25 is highly heterogeneous.
\end{itemize}

\section*{Acknowledgments}
The work at PRL is supported by the Dept. of Space, Govt. of India. We acknowledge the local staff at the Mount Abu Infra-Red Observatory for their help and a special thanks to Mr. Prashant Chauhan, for his assistance in the observations. We also thank our colleagues in the Astronomy \& Astrophysics division at PRL for their comments and suggestions. MH would like to thank the Slovak Grant Agency for Science VEGA, Grant No. 2/0023/18, and project ITMS No. 26220120029, based on the supporting operational Research and development program financed from the European Regional Development Fund. We would also like to thank the referee for constructive comments and suggestions.

\newpage

\bibliographystyle{aasjournal}


\label{lastpage}

\end{document}